# Free-Space Data-Carrying Bendable Light Communications

Long Zhu, Andong Wang, Jian Wang*

*Wuhan National Laboratory for Optoelectronics, School of Optical and Electronic Information, Huazhong University of Science and Technology, Wuhan 430074, Hubei, China.*

*Correspondence to:   jwang@hust.edu.cn*

**Abstract**: Bendable light beams have recently seen tremendous applications in optical manipulation, optical imaging, optical routing, micromachining, plasma generation and nonlinear optics. By exploiting curved light beams instead of traditional Gaussian beam for line-of-sight light communications, here we propose and demonstrate the viability of free-space data-carrying bendable light communications along arbitrary trajectories with multiple functionalities. By employing 39.06-Gbit/s 32-ary quadrature amplitude modulation (32-QAM) discrete multi-tone (DMT) signal, we demonstrate free-space bendable light intensity modulated direct detection (IM-DD) communication system under 3 different curved light paths. Moreover, we characterize multiple functionalities of free-space bendable light communications, including bypass obstructions transmission, self-healing transmission, self-broken trajectory transmission, and movable multi-receiver transmission. The observed results indicate that bendable light beams can make free-space optical communications more flexible, more robust and more multifunctional. The demonstrations may open a door to explore more special light beams enabling advanced free-space light communications with enhanced flexibility, robustness and functionality.

In recent years, bendable light beams have received a great deal of attention because of their tremendous application potential in a diversity of fields such as optical manipulation [1-3], optical imaging [4,5], optical routing [6], micromachining [7], plasma generation [8] and nonlinear optics [9-12]. Bendable light beams are a novel class of electromagnetic wave associated with a localized intensity maximum that propagates along a curved trajectory. Since the initial work studying Airy beam solutions of the paraxial wave equation propagating along parabolic trajectories [13-15], more general classes of bendable light beams have been reported including Mathieu beams along elliptical trajectories and Weber beams along parabolic trajectories [16-18]. Airy beam is one type of nondiffracting beams, which can maintain its wavefront during transmission just like Bessel beam. In contradistinction with the Bessel beam, the Airy beam possesses the properties of self-acceleration in addition to nondiffraction and self-healing, which propagates along a parabolic trajectory. In addition to Airy beam, bendable light beams can also reconstruct their wavefront, and continue propagate along the preset trajectory. To exploit the advantage of bendable light beams for different applications, one need to bend the light along arbitrary trajectories. An efficient way to realize arbitrary curved light beam is based on the caustic method, which associates the desired trajectory with an optical light caustic. This method can be implemented in both real space and Fourier space [19-24].

To take full advantage of curved light beams, we can also employ them in free-space optical communications. In traditional free-space optical communications, the optical path is always a straight line connecting the transmitter and receiver. However, there are sometimes obstructions between the transmitter and receiver, which makes communication failure. By using curved light beams for free-space optical communications, one can easily navigate around the obstructions by setting appropriate trajectories. Besides, bendable light beams are nondiffracting beams. When coming through an obstruction, they can reconstruct their wavefront and continue to propagate along the preset trajectories. Moreover, by designing proper trajectories, one can send information to multiple users and avoid unwanted users.

Thus, by employing bendable light beams in free-space optical communications, it will make the communication system more flexible and robust.

By simply using the phase only spatial light modulation, we realize bendable light beams along arbitrary trajectories including self-broken trajectory. By employing 39.06-Gbit/s 32-ary quadrature amplitude modulation (32-QAM) discrete multi-tone (DMT) signal, we successfully demonstrate free-space bendable light intensity modulated direct detection (IM-DD) communication system using 3 different curved light paths. Moreover, we also characterize the transmission performance under 4 different conditions: (i) bypass the obstructions communication; (ii) self-healing communication; (iii) self-broken trajectory communication; (iv) movable multi-receiver communication.

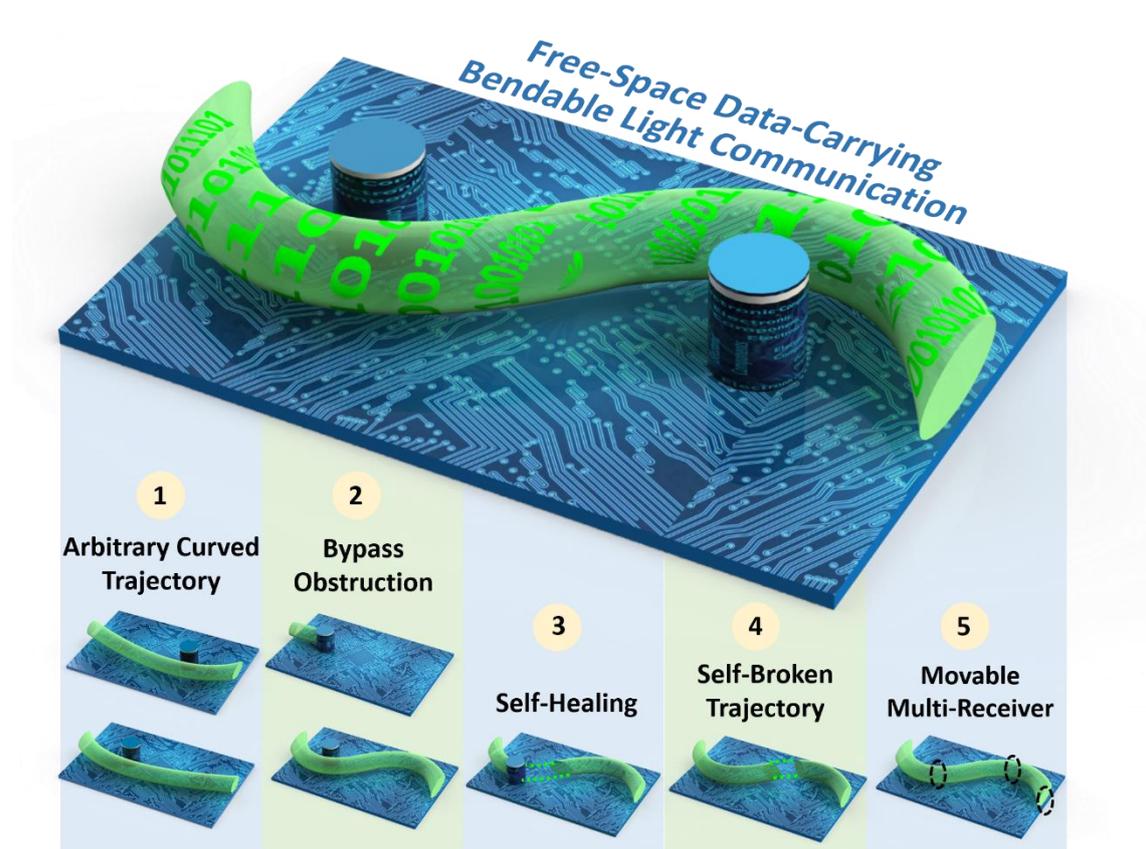

**Fig. 1.** Concept and principle of free-space bendable light communications.

Figure 1 illustrates the concept and principle of free-space bendable light communications. Here shows one example of free-space data-carrying bendable light communication system from one side of a circuit board to the other side. Firstly, by carefully designing the specific phase pattern for spatial light modulation via optical light caustic method (see supplementary materials), one can build arbitrarily curved light paths, which makes the communication system much more flexible. Secondly, in contrast to traditional Gaussian beam, the generated bendable light can also avoid or bypass existed obstructions as expected. Thirdly, the data-carrying bendable light can recover its wavefront when directly passing through the obstruction. The self-healing property of the curved light beam makes the communication system more robust. Fourthly, one can even construct a self-broken trajectory curved light beam, which can avoid unwanted users. Finally, owing to the self-healing property, the curved light can deliver the information to moveable multi-users along the curved light path. Consequently, by employing bendable light, the free-space communication system will become more multifunctional, more flexible and more robust.

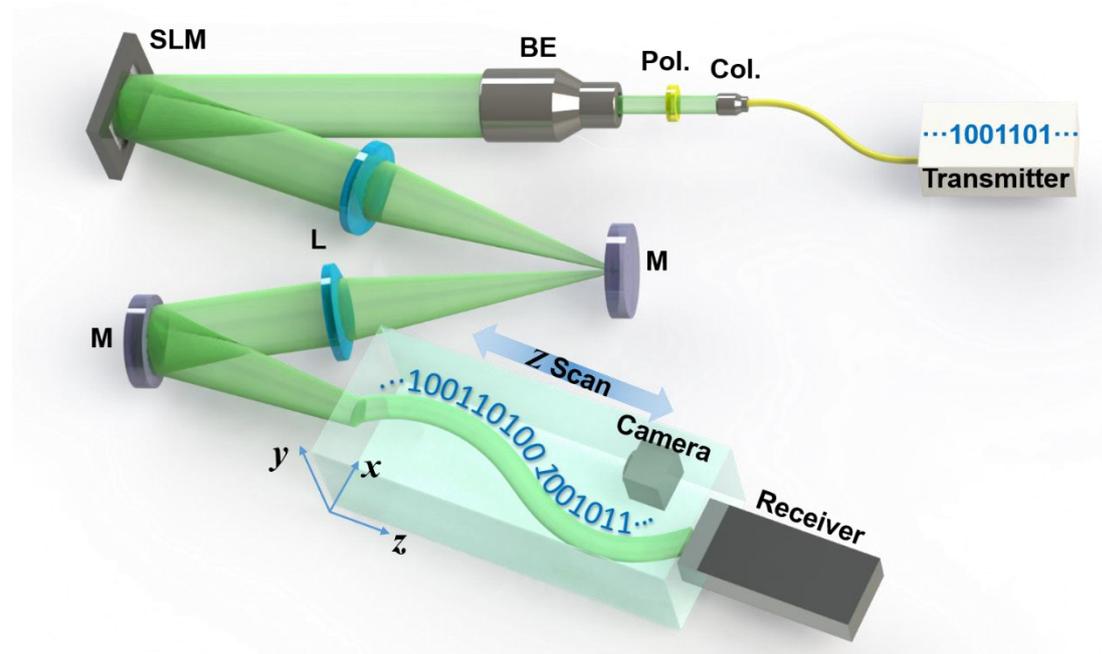

**Fig. 2.** Experimental configuration of free-space bendable light communications. Col.: collimator; Pol.: polarizer; BE: beam expander; SLM: spatial light modulator; M: mirror; L: lens.

The experimental configuration used in proof-of-concept demonstration of free-space bendable light communications is shown in Fig. 2 (see supplementary materials). A

39.06-Gbit/s 32-QAM DMT signal at 1550 nm from the transmitter is sent to the collimator to generate a free-space Gaussian beam. A polarizer (Pol.) is used for light polarization alignment with the polarization-sensitive spatial light modulator (SLM). Then the light is expended by a 5x beam expender (BE), which can illuminate the full extent of the SLM. The data-carrying bendable light is generated immediately after the SLM, which is loaded with the desired phase pattern by optical light caustic method for bendable light beam generation (see supplementary materials). In order to record the full propagating trajectory, a two lens 4f imaging system is employed. A camera is placed after the 4f system to record the propagation dynamics of the bendable light by moving along a motorized linear translation stages. At last, the curved light beam is coupled into the receiver for signal detection.

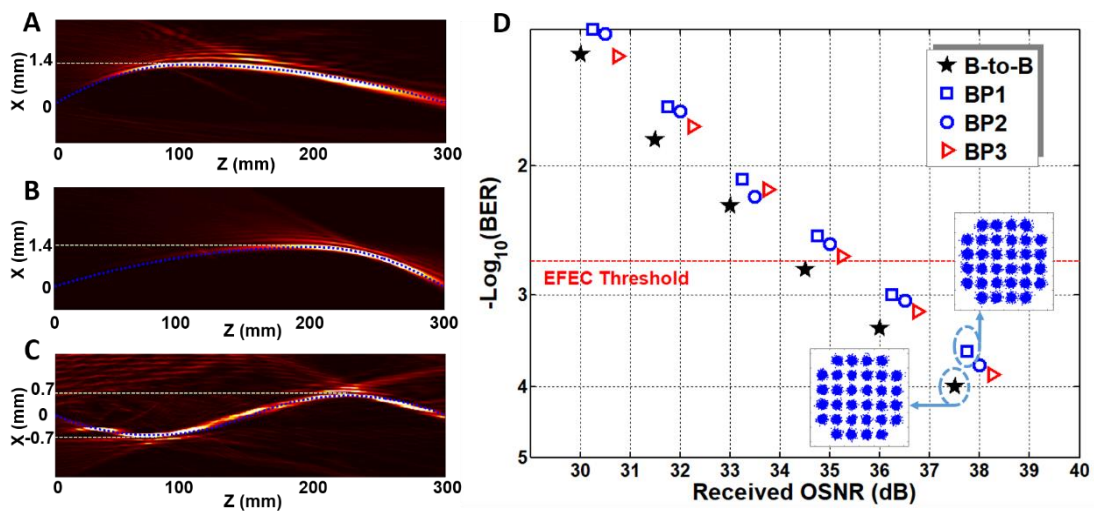

**Fig. 3.** Experimental results of free-space bendable light communications along arbitrary trajectories. (A), (B) and (C) are the measured intensity distribution of three different bendable light beams at *x-z* plane (the blue dashed line is the preset trajectory). (D) Measured bit-error rate (BER) performance of the three different data-carrying bendable light beams. Insets show constellations of 32-QAM DMT signals. B-to-B: back to back. BP1-BP3 correspond to (A)-(C). EFEC: enhanced forward error correction.

We first demonstrate free-space bendable light communications along arbitrary trajectories. Three bendable light beams with different curved trajectories are successfully generated. The measured intensity distributions are depicted in Fig. 3A, 3B and 3C. The propagating distance of the curved light beams are all 300 mm along the *z* direction. The curved light beams (BP1 and BP2) in Fig. 3A and 3B are along parabolic trajectories. The bending offset of them are both 1.4 mm. Moreover, we also generate S-shaped curved light beam (BP3), which is displayed in Fig. 3C. The bending offset of two peaks are both 0.7 mm. From the measured intensity distributions, one can clearly find that the measured bendable light beams are in good agreement with the predesigned trajectories as marked by blue dashed lines shown in Fig. 3A, 3B and 3C. Furthermore, we measure the bit-error rate (BER) performance as a function of the received optical signal-to-noise ratio (OSNR) for the three bendable light beams, as depicted in Fig. 3D. 39.06-Gbit/s 32-QAM DMT signals are employed in the IM-DD free-space communication system. The observed OSNR penalties at a BER of $2\times10^{-3}$ (enhanced forward error correction (EFEC) threshold) for the three bendable light beams (BP1, BP2 and BP3) are ~0.9 dB. The insets in Fig. 3D plot constellations of 32-QAM DMT signals.

We further demonstrate multiple functionalities of free-space bendable light communications. Firstly, we set obstructions along the line of sight between the transmitter and receiver, as shown in Fig. 4A. The Gaussian beam is also considered for comparison. One obstruction (Ob1) is set at *z*=75 mm and the other (Ob2) is set at *z*=225 mm. The diameter of the obstructions are both 0.8 mm. We first set one obstruction (Ob1), and measure the BER performance of the S-shaped light beam (curve BP-Ob-1). Then, we set two obstructions (Ob1 and Ob2) simultaneously, and measure the BER curve of the S-shaped light beam (curve BP-Ob-2). The BER performance of both conditions are almost the same as the one without obstructions (curve BP). The observed OSNR penalties at a BER of $2\times10^{-3}$ are about 0.9 dB. However, for Gaussian beam with obstructions, one cannot receive enough optical power for signal detection. The BER of Gaussian beam transmission with obstructions (curve Gauss-Ob) is about 0.5, as shown in Fig. 4A.

Secondly, we demonstrate the self-healing property of the free-space bendable light communications. An obstruction with a diameter of 0.8 mm is set at the curve path of the S-shaped bendable beam (*z*=75 mm), as depicted in Fig. 4B. Thus, the curved light is blocked by the obstruction. After propagation, the light reconstructs is wavefront. A receiver is placed at z=300 mm to receive the self-healing data-carrying bendable light. We also measure the transmission performance of the reconstructed light (curve BP-Re), as shown in Fig. 4B. As seen from the BER curve, one can easily find that the reconstructed light has almost the same performance with the non-blocked one (curve BP).

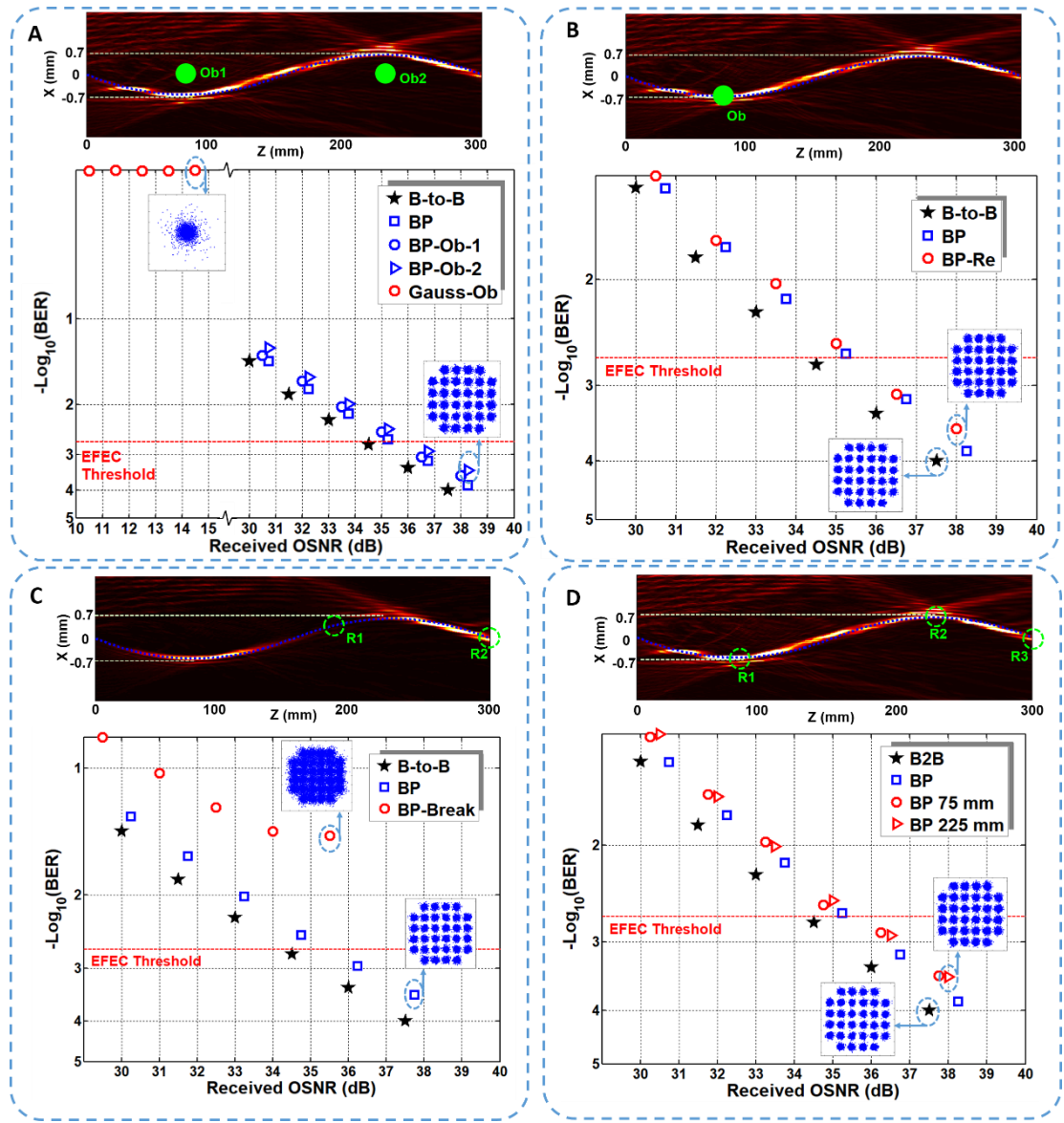

**Fig. 4.** Experimental results of free-space bendable light communications for different functionalities. (A) Measured intensity distribution and BER performance of the bendable light communication under obstruction condition. (B) Measured intensity distribution and BER performance of the bendable light communication under self-healing condition. (C) Measured intensity distribution and BER performance of the self-broken bendable light communication. (D) Measured intensity distribution and BER performance of the bendable light communication for movable multiple users.

Thirdly, we also generate a curved light beam with self-broken trajectory, as shown in Fig. 4C. From z=170 mm to z=200 mm, the main lobe of the curved light is missing, and then recovered at the end. Thus, one cannot detect the information at the broken part (from z=170 mm to z=200 mm). We then set the receiver at z=185 mm (R1), and measure the BER performance, which is shown in the BER curve (curve BP-Break). The received BER cannot reach the EFEC Threshold. Moreover, we also measure the BER performance at z=300 mm (R2), as shown in the BER curve (curve BP). The observed OSNR penalties at a BER of $2\times10^{-3}$ are about 1 dB, which means one can successfully receive the information at the end of the bendable light beam.

At last, we characterize the communication performance of the bendable light beam for multiple users. Owing to the self-healing property, the curved light can deliver the information to movable multiple users along the curved light path trajectory, which is not available for traditional free-space light communications. Here, we set three receivers along the light path (z=75 mm (R1), 225 mm (R2), and 300 mm (R3)), which is marked in Fig. 4D. The measured BER performance is plotted in Fig. 4D. The three receivers have almost the same transmission performance. The observed OSNR penalties at a BER of $2\times10^{-3}$ are about 0.9 dB.

In summary, we successfully demonstrate free-space data-carrying bendable light communications. Moreover, we characterize multiple functionalities of free-space bendable light communication, including bypass obstructions transmission, self-healing transmission, self-broken trajectory transmission, and movable multi-receiver transmission. The observed results indicate that bendable light can make the free-space optical communication more multifunctional, more flexible and more robust. We expect this scheme to be also scalable in propagation distance and bending offset. The demonstrations may open a door to explore more special light beams and facilitate more extensive free-space light communication applications with advanced flexibility, robustness and functionality.

by combined phase and amplitude modulation in the Fourier space, *Opt. Lett.* **38**, 3387-3389 (2013).